\documentclass[traditabstract]{aa}
\usepackage{epsfig}

\usepackage{graphicx,natbib}
\usepackage{natbib,subfigure} 
\usepackage{amsmath,amssymb}
\usepackage[T1]{fontenc}

\usepackage{txfonts}

\usepackage{hyperref}

\hypersetup{
   colorlinks=true,
   citecolor=blue,
   linkcolor=blue,
   urlcolor=black
   }

\usepackage[francais,english]{babel}

\def\brunt{Brunt-Väisälä }

\def\mearth{ M_\oplus}

\def\Rp{R_\mathrm{\,p}}
\def\Mp{M_\mathrm{\,p}}

\def\op{\omega_\mathrm{\,p}}

\def\Req{R_\mathrm{eq}}

\def\mcore{M_\mathrm{c}}

\def\hp{H_{\!P}}
\def\delt{\nabla_T}
\def\delad{\nabla_\mathrm{ad}}
\def\delmu{\nabla_\mu}
\def\delZ{\nabla_Z}
\def\delrad{\nabla_\mathrm{rad}}
\def\Rzm{R_\rho^{-1}}
\def\Rmin{R_{\mathrm{min}}^{-1}}
\def\RIm{R_I^{-1}}
\def \deld{\nabla_\mathrm{d}}
\def\delmean{\langle\nabla_T\rangle}

\def\amu{\alpha_{\mu}}
\def\at{\alpha_T}
\def\deltat{\delta_T}

\def \kapt{\kappa_T}
\def\Nt{N_T}

\def\cp{c_P}

\def\mum{\bar{\mu}}

\def\NL{N_\mathrm{l}}
\def\Ras{Ra_\star}
\def\NuT{Nu_T}
\def\Cl{C_L}
\def\Zenv{Z_\mathrm{env}}
\def\alphac{\alpha_\mathrm{crit}}
\def\surabad{\varepsilon_T}

\def\surabd{\varepsilon_\mathrm{d}}

\def\fconv{F_\mathrm{conv}}

\def\fdiff{F_\mathrm{d}}
\def\ftot{F_\mathrm{tot}}

\def\fdad{F_\mathrm{d}^\mathrm{ad}}

\def\Vg{V_G}

\def\Vrot{V_\mathrm{rot}}

\def\d{\mathrm{d}}

\def\n{\mathrm{n}}

\newcommand{\balign}[1]{
\begin{align}
#1
\end{align}}

\newcommand{\fracl}[2]{\raisebox{0.4ex}{$#1$} / \raisebox{-0.7ex}{$#2$}}

\newcommand{\eq}[1]{Eq.\,(\ref{#1})}
\newcommand{\eqs}[2]{Eqs.\,(\ref{#1}) and (\ref{#2})}
\newcommand{\fig}[1]{Fig.\,\ref{#1}}

\newcommand{\sect}[1]{Sect.\,\ref{#1}}

\newcommand{\app}[1]{Appendix\,\ref{#1}}
\newcommand{\tab}[1]{Table\,\ref{#1}}

\newcommand{\ind}[2]{\ #1 _{\!\! \mathrm{\ #2}}}

\newcommand{\pd}[2]{\frac{\partial \!\! \ #1}{\partial \!\! \ #2}}

\newcommand{\pdc}[3]{\left. \frac{\partial \!\! \ #1}{\partial \!\! \ #2}\right|_{#3}}

\newcommand{\dd}[2]{\frac{\mathrm{d} \!\! \ #1}{\mathrm{d}\!\! \ #2}}

\titlerunning{A new vision on giant planet interiors}
\authorrunning{Leconte \& Chabrier}

\begin{document}

\title{A new vision on giant planet interiors:\\
the impact of double diffusive convection}

\author{J\'er\'emy Leconte\inst{1} \and Gilles Chabrier\inst{1,2} 
}

\institute{ \'{E}cole normale sup\'erieure de Lyon, CRAL (CNRS), 46 all\'ee d'Italie, 69007 Lyon,\\ Universit\'e de Lyon, France (jeremy.leconte, chabrier @ens-lyon.fr)
\and
School of Physics, University of Exeter, Exeter
}

\date{Received 20 February 2009}

\offprints{J. Leconte}

\abstract{
While conventional interior models for Jupiter and Saturn are based on the simplistic assumption of a solid core surrounded by a homogeneous gaseous envelope, we derive new models with an inhomogeneous distribution of heavy elements, i.e. a gradient of composition, within these planets. Such a compositional stratification hampers large scale convection which turns into double-diffusive convection, yielding an inner thermal profile which departs from the traditionally assumed adiabatic interior, affecting these planet heat content and cooling history. 

To address this problem, we develop an analytical approach of layered double-diffusive convection 
and apply this formalism to Solar System gaseous giant planet interiors. These models satisfy all observational constraints and yield a metal enrichment for our gaseous giants up to 30 to 60\% larger than previously thought. The 
models also constrain the size of the convective layers within the planets. As the heavy elements tend to be redistributed within the gaseous envelope, the models predict smaller than usual central cores inside Saturn and Jupiter, with possibly no core for this latter. 

These models open a new window and raise new challenges on our understanding of the internal structure of giant (solar and extrasolar) planets, in particular on the determination of their heavy material content, a key diagnostic for planet formation theories.
}




\maketitle

\section{Introduction}
\label{sec:intro}
More than 500 planets have now been discovered orbiting stars outside our Solar System, spanning a range from a few Earth masses to several Jupiter masses. Planets thus seem to be ubiquitous in nature. These discoveries raise fundamental questions about the inner composition, evolution and origin of these bodies, and about the fundamental properties of planets in general, including the ones of our own Solar System. Characterizing their internal structure and composition, and from there, better understanding planet formation is one of the major challenges of modern astronomy. The determination of the heavy element content, for instance, provides key constraints to planet formation models,
in particular on the efficiency of solid planetesimal accretion
in the protoplanetary disk to build a planet embryo. 
While only the mass of most of these planets can be derived from observations, the mean density can be inferred for about 20\% of these objects, as they transit their parent star, constraining the planet's gross composition (see \citealt{BCB10} for a recent review). Although providing an important diagnostic, however, this information is too limited to determine
 the element distribution within the planet and thus its precise compositional and thermal structure.
Assuming planet formation is a universal process, one thus must turn to our Solar System planets, in particular the two gas giants, Jupiter and Saturn, which encompass 92\% of the planetary mass of the Solar System, to derive more detailed interior models. Indeed, for our own giants, the gravitational moments have been determined with high accuracy by the various flyby missions entering their atmosphere and provide stringent constraints on their inner element distribution. 

An important question, for instance, is to determine whether the heavy elements present in giant planet interiors are located in a central core or are mixed in a large
fraction into the hydrogen-helium (H/He) fluid envelope. In this latter case, a major issue is to determine whether convective mixing is sufficiently efficient to yield a homogeneously mixed envelope or, alternatively,
if giant planet interiors can exhibit a \textit{continuous} compositional gradient. Exploring such a possibility is crucial  to determine (i) the maximum amount of heavy 
elements compatible with observational constraints, (ii) the efficiency of heat transport in giant planets. These two issues
directly impact the planet mechanical (density), chemical (composition) and thermal (luminosity, temperature) structures at a given age, with major consequences
on our understanding of planet formation and evolution.

Traditionally, giant planet models have always been based on two major specific assumptions concerning their inner profile,
essentially for reasons of simplicity in the planet's modeling (see e.g. \citealt{Ste85}). It is conventional to assume (i) that the inner structure of our giants consists of a few - generally 2 to 3
 - superposed, well separated, {\it homogeneous} regions, namely, going from the planet's center to the surface, a central solid rocky/icy core, and a surrounding largely
dominantly H/He gaseous envelope,
often split into an inner metallic region and an outer atomic/molecular one; the more dense components are always supposed to have been accreted first or to have quickly settled into the centre under the action of gravity (for planetesimals accreted after the runaway gas accretion); 
(ii) that planetary interiors are adiabatic, based on the fact that the giant planet heat flow must be transported by convection \citep{Hub68}. 
All the present determinations of the internal - chemical, mechanical and thermal - structures of the Solar system planets, including their heavy material content, are derived assuming such homogeneously stratified, adiabatic interiors (\citealt{SG04}; \citealt{FN10}).

Giant planet interiors, however, might depart from this conventional, simplified description, because of complex processes for which we lack an accurate description but which may very well be at play in real situations \citep{Ste85}.
In this paper, we derive interior models for Jupiter and Saturn which relax the aforementioned preconceptions. Instead of the homogeneous layer assumption, we explore the possibility of a mixed, {\it inhomogeneous} solid-gas interior composition, leading to a heavy material gradient throughout the planet. 
This in turn tends to suppress large scale convection which, due to the double diffusive instability (see \sect{sec:adiab}), can turn into either turbulent enhanced diffusion or  layered convection. As both these heat transport mechanisms are fairly inefficient compared to usual convection, this compositional gradient thus leads to significant departure from global adiabaticity in the interior.
As shown below, these models do fulfill the planet observational constraints while leading to (i) a significantly larger metal content
and (ii) significantly larger internal temperatures  than the one inferred from homogeneously stratified adiabatic models. 

This opens a new vision on planet structure, evolution and formation efficiency. Such inhomogeneous interior profiles for Solar System giant planets had briefly been suggested several decades ago by Stevenson \citep{Ste85} but no attempt has ever been made to derive consistent models and to verify whether such models would be consistent with the planet various observational constraints. This scenario has been revived recently in the context of extrasolar planets and has been shown to provide a possible or at least complementary explanation for the anomalously large observed radii of many of these bodies \citep{CB07}. Indeed, as mentioned above, not only an inhomogeneously stratified interior yields a different interior structure and global metal content, but it decreases heat transport efficiency throughout the planet's interior and thus affects its cooling, thus its mass-radius relationship at a given age, a crucial diagnostic to
understand (transiting) extrasolar planet structure and evolution. Since, as mentioned earlier, only loose constraints on the object's internal composition are accessible for gaseous exoplanets, it is crucial to verify whether such unconventional internal structures are a viable possibility for our own giants. Furthermore,
determining the maximum possible amount of heavy elements in Jupiter and Saturn and their distribution within the planet are important diagnostics to understand how our own Solar System
giants formed.

In order to address this issue, we first briefly review our current understanding of the double diffusive instability, and of the various regimes under which it can occur in \sect{sec:intro_semiconv}.
Then, in \sect{sec:mlt}, we derive an analytical formalism, based on a standard parametrization of convection, similar to the mixing length formalism, which describes the global transport properties of an inhomogeneous convective/diffusive medium. In \sect{sec:mlt_limit}, we discuss the possible existence of an equilibrium size for the convective/diffusive layers present in a semi-convective planet and derive analytical constraints for this equilibrium value. Finally, in \sect{sec:resnum}, we derive semi-convective models of Jupiter and Saturn that are consistent with the gravitational moments and with the surface abundances measured by the Galileo and Cassini missions. This enables us to further constrain the number of possible convective/diffusive cells within these planets, and to derive new values of the heavy element content and of the core masses for our gas giants.


\section {Physical mechanisms leading to inhomogeneous density stratification}\label{sec:intro_semiconv}

\subsection{Compositional gradient}

In  the present calculations, in contrast to all previous planetary models, we consider a mass fraction of heavy material $Z(m)$ at a depth $r(m)$ within the planet (i.e. at the
depth of the iso-density surface enclosing a mass $m$ of the planet) that is \textit{continuously} decreasing from the core to the surface, producing a compositional gradient within the gaseous envelope of the planet, \balign{\delZ \equiv\dd{\ln Z}{\ln P }=\dd{\,r}{\ln P}\frac{1}{Z}\dd{\,Z}{\,r}\equiv -\frac{\hp}{Z} \nabla Z, \label{gradientz}}
where $\hp\equiv-\dd{\,r}{\ln P}$ is the pressure scale height.

The possible origin of such a compositional gradient is an open question 
\citep{Ste85,CB07}.
In the conventional scenario, all the accreted planetesimals are assumed to directly sink
to the core and not to evaporate in the envelope, for sake of simplicity. In reality, however, incomplete mixing of large planetesimals or dissolution of a substantial fraction of volatiles and rocks from small
solid bodies could occur in the envelope during the phase of planetesimal accretion on the nascent planet; a substantial amount of ice could as well remain in the envelope \citep{IP07,HI11}.

The gradient might also stem from an only partial redistribution by small scale convective motions of stably layered (soluble) constituents released by core erosion in the gas-rich envelope during the planet's evolution \citep{Ste82,GSH04}, as seems to be supported for water by recent numerical simulations \citep{WM10}.
This could be enhanced by the immiscibility (phase separation) of an abundant enough material (e.g. helium, water) in the dominantly metallic-hydrogen envelope \citep{SS77b}. This would change the dynamical properties of (double diffusive) convection near the regions of immiscibility and add complexity to the problem. Therefore, to avoid extra complication, we only consider the occurence of this process for soluble material in the present paper.

At last, rapid rotation and/or strong magnetic fields, necessarily present in Jupiter and Saturn interiors%
, are known to hamper large-scale convection \citep{CGB07}, possibly leading to imperfect
mixing of heavy elements in part of the envelope.




\subsection{Double diffusive convection}\label{sec:adiab}

\begin{figure*}[htbp] 
  \sidecaption
\resizebox{.7\hsize}{!}{\includegraphics{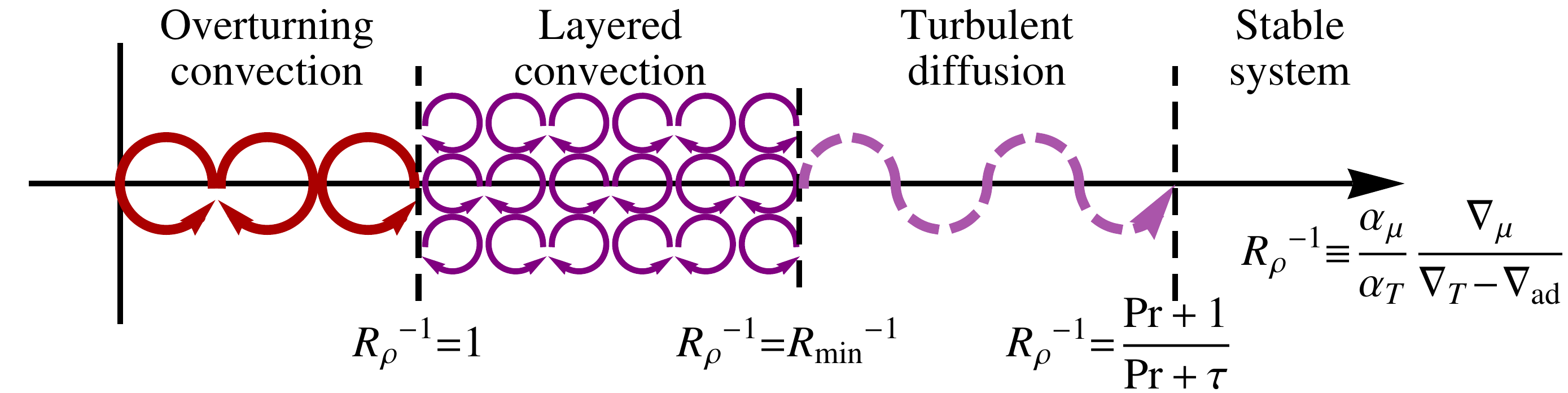}}
 \caption{
Stability diagram for a medium presenting a destabilizing temperature gradient and a stabilizing solute gradient, as a function of increasing $\Rzm$ (decreasing $R_\rho$). The usual instability Ledoux criterion corresponds to $\Rzm\leqslant1$.}
 \label{fig:stab_diagram}
\end{figure*}

Convective systems in which (rapidly diffusive) heat and (slowly diffusive) composition have opposite destabilizing and/or stabilizing effects tend to develop inhomogeneous density profiles.
The (de)stabilizing effect of heat can be quantified by the thermal gradient, $\delt\equiv\dd{\ln T}{\ln P}$, or more precisely by the super adiabaticity ($\delt-\delad$, where the derivative in $\delad$ is taken at constant specific entropy), and the one of the heavy elements by the gradient of mean molecular weight ($\mu$), $\delmu\equiv\dd{\ln \mu}{\ln P}$. 
The actual dynamical state of the medium depends on the value of the \textit{density ratio}
\balign{R_\rho\equiv\frac{\at}{\amu}\,\frac{\delt-\delad}{\delmu},}
where $\at\equiv -\pdc{\ln \rho}{\ln T}{P,\mum}$, $\amu \equiv \pdc{\ln \rho}{\ln \mum}{P,T}$ \footnote{For a perfect gas, $\amu=1$ and $\at=1$, as will be used hereafter.} \citep{Ste60}.

In the case both the mean molecular weight (due e.g. to a higher concentration of salt in salty water) and the temperature increase with height ($\delmu$ and $(\delt-\delad)<0$), the compositional gradient is destabilizing while the temperature gradient is stabilizing. This case is referred to as the \textit{fingering case}. In that case, convective instability develops when $R_\rho<1$, which is equivalent to the \textit{Ledoux instability criterion}. But even if $1<R_\rho<1/\tau$, where $\tau= D/\kapt$ is the ratio of solute ($D$) to thermal ($\kapt$) diffusivities, the slower diffusivity of elements compared to heat yields the so-called \textit{double-diffusive} instability which leads to the formation of salt fingers, and sometimes of thermo-compositional staircases, as observed in some parts of the oceans and in laboratory experiments \citep{Tur67}.

The opposite case, referred to as the \textit{diffusive case}, corresponds to a fluid exhibiting a \textit{positive} molecular weight gradient ($\delmu>0$).  In that case, the fluid will be convectively unstable if this gradient is insufficient to stabilize the system
against convective instability, i.e. if the Ledoux \textit{instability} criterion
\balign{\delt-\delad>\frac{\amu}{\at}\,\delmu\,\Leftrightarrow\,R_\rho>1,\,\,{\rm i.e.}\,\,\Rzm<1,}
is met\footnote{As discussed by \citet{RGT11}, the analogy between the fingering and the diffusive case is more apparent when the \textit{inverse} density ratio, $\Rzm$, is used, as will be done thoughout the rest of the paper.}. What happens, however, in the regions which are stable according to the Ledoux criterion, i.e. $\Rzm>1$, but unstable according to the Schwarzschild criterion, 
\balign{\delt>\delad,} is less clear, especially in the astrophysical context, where the very low values of the Prandlt number ($Pr\equiv\nu/\kapt$, where $\nu$ is the kinematic viscosity) render difficult direct numerical hydrodynamical simulations. In particular, the exact nature of double diffusive convection if it occurs - homogeneous oscillatory convection or layered convection (i.e. uniformly mixed convective layers separated by thin diffusive interfaces characterized
by a steep jump in the mean molecular weight) - remains uncertain. Analytical arguments \citep{Rad03} and recent 3D hydrodynamical simulations \citep{RGT11,MGS11}, however, seem to suggest the picture presented in \fig{fig:stab_diagram}. When the mean molecular weight gradient ($\propto \Rzm$) decreases in a stable medium, homogeneous oscillatory convection, also called turbulent diffusion, first appears for 
\balign{\Rmin \leqslant \Rzm \leqslant \frac{Pr+1}{Pr+\tau}\,\,\mathrm{(oscillatory \ convection),}\label{oscil_conv}} while well defined thermo/compositional layers start to develop when 
\balign{1 \leqslant \Rzm \leqslant\Rmin \,\,\mathrm{(layered\ convection)}.\label{layerform}}
$\Rmin$ corresponds to the point where the solute to heat buoyancy flux ratio ($\equiv\gamma^{-1}$) stops decreasing when $\Rzm$ increases (see \citealt{Rad03} for details).
Its exact value, however, depends on the characteristics of the medium in a non trivial way and is difficult to estimate \citep{RGT11,MGS11}. For smaller ${\delmu}$ gradients, the medium is unstable according to the Ledoux criterion, and the thermal forcing is strong enough to force large scale overturning convection.

Various arguments seem to support, or at least not to exclude, the existence of layered convection under planetary conditions \citep{CB07}. Conducting 3D hydrodynamics calculations over a wide domain of parameter space (Prandlt number and atomic to thermal diffusivity ratio) {\it including the regime relevant for planetary interiors}, \citet{MGS11} always find a domain where $\gamma^{-1}$ decreases with $\Rzm$, a necessary and sufficient condition for the layering instability, thus layer formation to occur. A central question is then the size of the layers; this will be
examined in \sect{sec:mlt_limit}.
In any event, both homogeneous double-diffusive convection or layered convection - generically denominated as "semi-convection" in the following - are found to yield thermal and compositional fluxes that are significantly smaller than that expected from standard convection. Indeed, the presence of diffusive interfaces
strongly decreases the efficiency of heat transport compared with large-scale, adiabatic convection, leading in planet interiors to a significant departure from the usual adiabatic profile, as quantified below.



\section{An analytical theory for layered convection}\label{sec:mlt}

\begin{figure*}[tbp] 
 \centering
 \resizebox{1.\hsize}{!}{\includegraphics{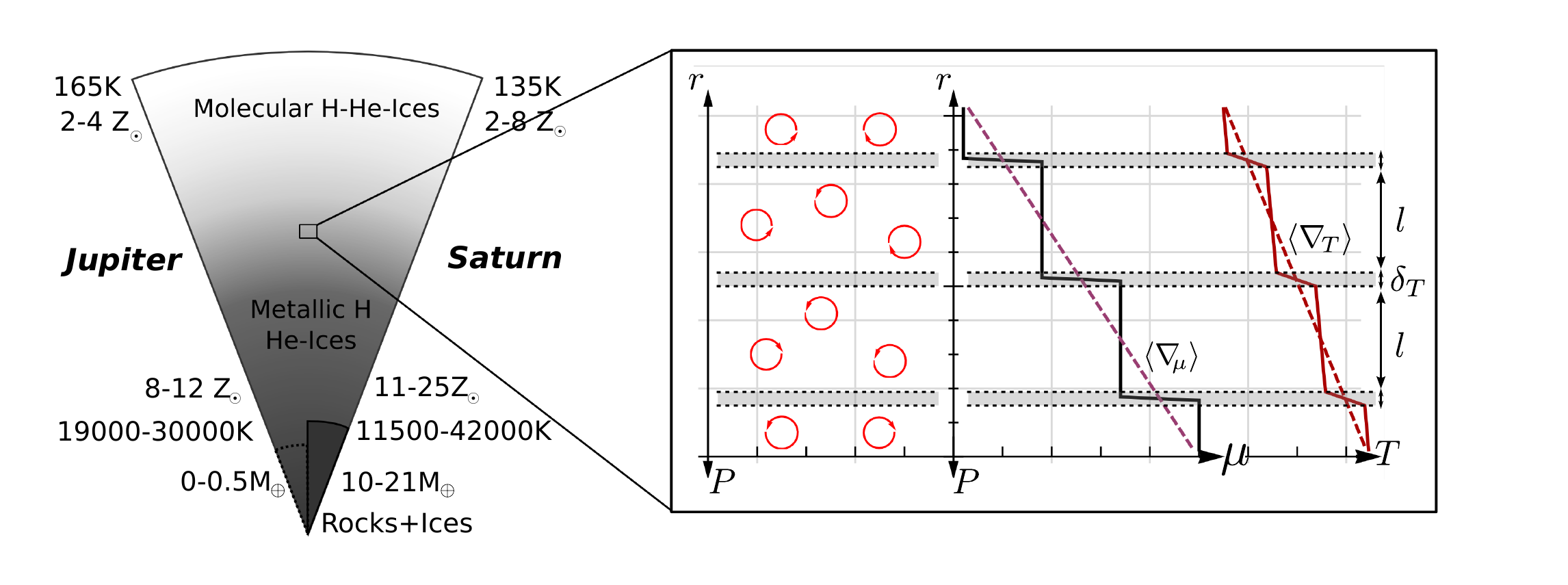}}
 \caption{
Schematic representation of the interiors of Jupiter and Saturn, according to the present study, and of layered convection, with the resulting compositional and thermal radial profiles. The abundance of metals is constant within the well mixed convective cells of size $l$, and undergoes a steep variation within the diffusive interfaces of thermal size $\delta_T$ (grey regions). Thanks to this steep gradient, these interfaces are stable against convection and energy and matter are transported therein by diffusive processes. Because the size of these layers is very small compared with the size of the planet, the mean thermal and compositional gradients ($\delmean$ and $\langle\delmu\rangle$) can be used in good approximation to infer the global planet structure.}
 \label{fig:cartoon}
\end{figure*}

In order to investigate the impact of such strongly hampered convection on giant planet internal structure, we developed a simple sub-grid model of layered convection. As illustrated on \fig{fig:cartoon} and found in simulations \citep{RGT11,MGS11}, we consider that a semi convective zone consists of a large number, $\NL$, of well mixed convectively unstable layers of size $l$, separated by thin diffusive interfaces of {\it thermal} thickness $\delta_T$, within which the large stabilizing compositional gradient completely inhibits convective motions. 

\subsection{Convective layers}\label{sec:conv_layers}

Within each convective layer, the fluid is expected to follow the dynamics found in turbulent \textit{Rayleigh-Bénard} convection within a cell of typical height equal to the size of the layer, $l$. By analogy with the mixing length formalism (see details in \app{sec:mlt_app}), we define a dimensionless mixing length parameter by dividing $l$ by the pressure scale height, $\alpha\equiv l/\hp$. 

In a laboratory or a numerical experiment, the efficiency of the convection is characterized by the thermal \textit{Nusselt} number, 
\balign{\NuT\equiv\frac{\ftot-\fdad}{\fdiff-\fdad},} where by definition the total \textit{intrinsic} flux ($\ftot$), the flux transported by diffusive processes ($\fdiff$), and the diffusive flux that would be present in a completely adiabatic zone ($\fdad$) are given by \citep{CG68,HK94}
\balign{\left(\begin{array}{c}\ftot \\\fdiff \\ \fdad\end{array}\right)\equiv\kapt\frac{\rho\,\cp T }{\hp}\left(\begin{array}{c}\deld \\ \delt \\ \delad\end{array}\right) ,}
where $\cp$ is the heat capacity at constant pressure. 
It is found that, for large Rayleigh numbers, the Nusselt number follows a law of the type 
\balign{\NuT=\Cl\,\Ras^a,\label{nuras}}
where $\Ras$ is a \textit{modified} \textit{Rayleigh} number\footnote{Indeed, in the astrophysical context, it is convenient to use $\Ras=Pr \times Ra$, where $Ra$ is the usual Rayleigh number.}, which is the ratio of the strength of the thermal forcing to the one of the radiative losses
\balign{\Ras\equiv \frac{\alpha_T\, g \hp^3}{\kapt^2} \,\alpha^4\,(\delt-\delad)= \left(\fracl{\Nt^2 l^4}{\kapt^2}\right). \label{def_rastar}}
Here, $g$ is the local gravity acceleration, and $\Nt$ the \brunt frequency.


As convection at very high Rayleigh numbers is difficult to study either experimentally or computationally, it is difficult to give precise values for $a$ and $\Cl$. For the bounded Rayleigh-B\'enard problem, theoretical models suggest that the exponent of the convective flux law, $a$, could be equal to 1/3 \citep{GOM10}\footnote{Although the simulations presented by \citet{RGT11} seem to support exponent values smaller than 1/3, suggesting that interfaces act as impermeable boundaries, it should be noticed that the height of the layers present in their simulations is small compared to a pressure scale height. Their conclusion may thus not be valid for larger layers.
}. However, experiments done by \citet{Kri95} tend to show that this exponent could be smaller, and as low as $a=0.2$. On the other hand, for homogeneous Rayleigh-Bénard convection (without boundaries), \citet{GOM10} showed that the regime predicted by the mixing length theory, i.e. $\NuT=\Ras^{1/2}$ ($\Cl=1$, $a=1/2$; see \app{sec:mlt_app}), is recovered. In the following, we will thus consider $0.2\leqslant a \leqslant 0.5$ and $\Cl=1$.

\eq{nuras} is sufficient to calculate the flux transported by convection once the super adiabaticity is known. To compute this latter, however, we must first define a quantity which can be computed \textit{a priori} from the local thermodynamical properties of the medium and the total internal energy flux to be transported. Following \citet{HK94}, this \textit{convective forcing} can be defined by 
\balign{\label{def_Phi}
\Phi&\equiv \NuT\times\Ras.
}
Introducing $\surabd\equiv\deld-\delad,$ we rewrite \eq{def_Phi} as $\Phi\equiv \Phi_0\,\alpha^4\, \surabd$ where
\balign{\Phi_0\equiv  \left(\frac{\alpha_T g \hp^3}{\kapt^2}\right).}
It is clear from \eq{def_Phi} that $\Phi_0$ is a local constant of the medium, which characterizes its ability to transport energy by convection, independently of the mixing length or of the flux to be transported ($\propto \surabd$).

Then, from \eqs{nuras}{def_Phi}, one sees that in a region where convection remains efficient enough,
\balign{\label{solRas}
\Phi=\NuT\times\Ras=\Cl\,\Ras^{1+a}\,\,\Rightarrow\,\,\Ras=\left(\frac{\Phi}{\Cl}\right)^{1/(1+a)},
}
which yields the super adiabaticity,
\balign{\surabad\equiv\delt-\delad=\left(\frac{\surabd}{\NuT}\right)=\left(\frac{\surabd}{\Cl \,\Phi_0^a \,\alpha^{4\,a}}\right)^{1/(1+a)}.\label{surabad_layer}}
The range of super adiabaticity in the convective layers implied by this equation for the various possible exponents $a$ is shown in \fig{fig:surabad} (pale red area). As seen, the uncertainty on $a$ leads to a large dispersion on this super adiabaticity.
In this high convective efficiency regime, we can further compute the mean convective flux which, by definition, is given by
\balign{\fconv=\kapt \frac{\rho\, \cp T}{\hp}\, (\delt-\delad) \times \NuT. \label{def_fconv}}

As mentioned above, this scaling law only applies to the vigorous convection regime, i.e $\Phi$ or $\Ras \gg 1$. Thus, in terms of the layer height, convection remains efficient as long as $\alpha\gg\alphac$, where $\alphac$ is defined such that $\Ras(\alphac)=1$, and, from \eq{def_rastar}, the critical layer size is 
\balign{l_\mathrm{crit} =\left(\fracl{ \kapt ^2}{\Nt^2}\right)^{1/4}=d \,Pr^{-1/4},}
where $d$ is the lengthscale of the fastest growing mode of the linear instability, $d\equiv(\nu \kapt/\Nt^2)^{1/4}$ (e.g. \citealt{BG69}).
Analytical and numerical arguments show that the size of the fastest growing layers is equal to 10-100$\,d\gtrsim l_\mathrm{crit}$ \citep{Rad03,RGT11}. The efficient convection regime is thus appropriate 
in the planetary domain, where $Pr\sim10^{-2}-10^{-1}$ \citep{CB07}.

For conditions prevailing in the interior of the actual Jupiter (Saturn), the mean thermal diffusivity is $\kapt\sim5\times 10^{-5}$\,m$^2$.s$^{-1}$ \citep{Pot99}, $\Phi_0$ is equal to $\Phi_0\approx 3\times 10^{33}$ (9$\times 10^{32}$) and $\surabd\approx$\,10 (10; see also \fig{fig:thermalgradient}), so that $\alphac=2\times 10^{-9}$ (3$\times 10^{-9}$). In the following, all the order of magnitude estimates done throughout the text will use these values.

\subsection{Interfaces}\label{sec:interfaces}

In the interfaces of thermal size $\delta_T$, overturning convection is inhibited by the strong jump in molecular weight. However, these regions do not need to be in the fully diffusive regime ($\Rzm>(Pr+1)/(Pr+\tau)$), but can also be in the oscillatory convection regime characterized by \eq{oscil_conv}. Indeed, as shown by \citet{Rad05} the condition $\RIm>\Rmin$, where $\RIm$ is the inverse density ratio within the interface, is a sufficient criterion not only to ensure the stability of the interface itself, but also of the whole stack of layers which would otherwise merge into one homogeneous layer (see \sect{sec:merging_equilibrium} for details).

Therefore, the interface is very likely either in a stable diffusion state, or in a state of weakly turbulence enhanced diffusion. \citet{RGT11} show that in this regime $\NuT\lesssim2$, meaning that most of the energy is transported by pure diffusion. We will thus assume that the thermal gradient to be used in these regions is given by the gradient needed to transport the whole outgoing energy flux by diffusion: \balign{\label{def_deld}\deld\equiv\frac{1}{\kapt}\frac{\hp}{\rho\,\cp T }\ftot.}
The thermal diffusivity $\kapt$ encompasses the contribution of all diffusive processes. If diffusion is ensured by photons, as generally in most astrophysical objects,
the diffusive thermal gradient reduces to the so-called radiative gradient, $\delrad$ (e.g. \citealt{HK94}). However, in the deep interior of giant planets and in degenerate bodies, density can be high enough for the electrons to become degenerate enough to efficiently conduct thermal energy (see \citet{SS77a} and \citet{CB07} for the characteristic radiative and conductive opacities under jovian planet conditions). Along this paper, we will use the generic denomination $\deld$ for the diffusive temperature gradient, keeping in mind that diffusion is now due to electronic or atomic motions, with a characteristic thermal diffusivity $\kapt$. In the present calculations, we use the conductive thermal diffusivities calculated by \citet{Pot99}. 

\subsection{Mean thermal gradient}\label{sec:mean_T_grad}

Once we have calculated the thermal gradient in the convective zones of size $l$, and in the diffusive interfaces of size $\delta_T$, we need to determine
 the \textit{mean} properties of a whole stack of convective-diffusive cells. As discussed in \citet{CB07}, since the convective plumes must be fed by the diffusive interfaces, the thermal convective ($\Nt^{-1}$) and diffusive ($\deltat^2/\kapt$) time scales should be similar in each respective layer. Therefore
\balign{\label{del_over_l}
\left(\fracl{\deltat}{l}\right) =\left(\fracl{\kapt}{l^2 \,\Nt}\right)^{1/2}=\Ras^{-1/4}=\left(\fracl{\Phi}{\Cl}\right)^{-\frac{1}{4\,(1+a)}}.
}
Not surprisingly, the size of the interface, $\deltat$, is related to the lengthscale of the most unstable mode of the linear instability by $\deltat=~d/Pr^{1/4}.$
In addition, comparing the mean kinetic energy of an upwelling eddy with the potential energy barrier created by the negative buoyancy in the diffusive interface, we see that the above condition also entails that convective overshooting can be neglected (see \citealt{CB07} for details).

\begin{figure}[htbp] 
 \centering
 \resizebox{.9\hsize}{!}{\includegraphics{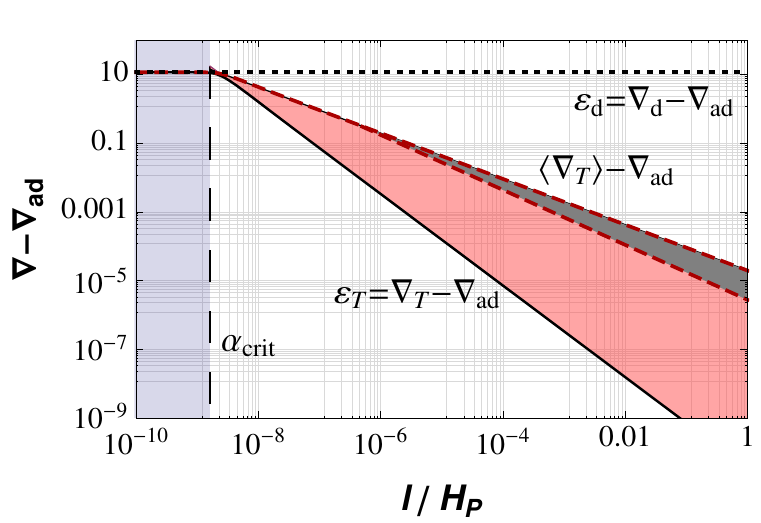}}
 \caption{Range of mean super adiabaticity ($\delmean-\delad$) of a semi-convective medium as a function of the mixing length parameter $\alpha=l/\hp$, for $0.2\leqslant a \leqslant 0.5$ (gray area between the red dashed curves; $\Phi_0=3\times 10^{33}$; $\surabd=10$; $\Cl=1$). For comparison, the super adiabaticity in a convective (pale red area; same parameters) and in a diffusive cell (dotted curve) are also shown. As expected a smooth transition between the convective and diffusive regimes occurs near $\alphac$ (see \app{sec:mlt_app}; here $\alphac\approx 2\times10^{-9}$).}
 \label{fig:surabad}
\end{figure}

Knowing $\deltat/l$ enables us to compute the \textit{mean} thermal gradient to be used to compute the planet's structure (see \sect{sec:method})
\balign{
\delmean\equiv\frac{\deltat}{l+\deltat}\,\deld+ \frac{l}{l+\deltat}\,\delt. \label{mean_gradT}
}
Substituting $\delt$ and $\deltat/l$ by their expressions in \eqs{surabad_layer}{del_over_l}, and developing the mean gradient to first non vanishing order, we find that
\balign{
\delmean\approx\delad+ (\deld-\delad)\, \left(\frac{1}{\Ras^{1/4}}+\frac{1}{\NuT}\right)+\mathcal{O}\left(\frac{1}{\Ras^{1/2}},\frac{1}{\NuT\Ras^{1/4}}\right)
,\label{meanT}}
or equivalently
\balign{\delmean - \delad &\approx (\deld-\delad) \, \left[ \left(\fracl{\Phi}{\Cl}\right)^{-\frac{1}{4(1+a)}} + \left(\Phi\,\Cl ^{1/a}\right)^{-\frac{a}{(1+a)}} \right] . \label{meanT2}}
As expected, two terms appear in the mean super adiabaticity, which is shown in \fig{fig:surabad} (dark grey region). The first one is due to the temperature jumps at each interface and is controlled by the convective overturning timescale which determines the size of these interfaces. The second one is due to the super adiabaticity in the convective layers themselves, which is enhanced by their small size. The relative contributions of these two terms only depend on the properties of the convection, characterized by the exponent $a$.

 Interestingly, if the impact of $a$ on the super adiabaticity in the convective layers (pale red area in \fig{fig:surabad}) is significant, the impact on the \textit{mean} super adiabaticity of the whole stack of layers, $\delmean - \delad$ (dark gray region), is rather small. This can be understood as follows: for the smallest value of the exponent considered here, $a=0.2$, convection is very inefficient and most of the mean super adiabaticity is contained in the convective layers (upper contour of the grey region). In that case, \eq{mean_gradT} is dominated by the second term, $\delmean\approx \delt$.

When the exponent is increased, the super adiabaticity in the convective layers ($\propto \Phi^{-a/(1+a)}$) drops dramatically, but this entails a growth of the convective timescale, and thus of the thickness of the interfaces (see \eq{del_over_l}). For efficient convection, $a\geqslant1/4$, this effect eventually dominates, and the mean super adiabaticity increases again with $a$.
This yields a rather narrow region for the mean super-adiabaticity within the considered range of exponent variation.

 Interestingly, to first order, the $a=1/2$ and $a=0.2$ cases, i.e. the two extreme values considered here, yield the same mean superadiabaticity. {\it Therefore, the model presented here for semi-convection depends only weakly on the precise prescription chosen for convection}. In the following, we will show results obtained with the mixing length theory, i.e. $\Cl=1$ and $a=1/2$, while keeping in mind that for $\alpha\gg\alphac$ these results would be the same as the one obtained in the $a=0.2$ case.

\subsection{Mean solute gradient and flux}

Disregarding convective overshoot, the solute flux ($F_Z$) is determined by the transport at each interface. As for heat, the turbulent transport of solute is expected to be very inefficient, and we will assume purely diffusive processes (see \sect{sec:interfaces}). The impact of a more efficient transport will be discussed in the next section. 
In this regime, the solute flux is thus given by
\balign{\label{flux_Z}
F_Z=-\rho\, D\, \nabla Z_{\,I},
}
where the subscript $I$ describes quantities determined within the interface, indicating that the solute flux $F_Z$ is determined by the element fraction variation rate at each interface. Neglecting the small inhomogeneities in the convective layer, the interfacial and mean gradients are related by
\balign{\nabla_{\mu,I}=\frac{l+\delta_Z}{\delta_Z}\langle\delmu\rangle \ \ \ \mathrm{or}\ \ \   \nabla Z_{\,I}=\frac{l+\delta_Z}{\delta_Z}\langle\nabla Z\rangle ,\label{dmu_dz_def1}}
where $\langle\nabla Z\rangle$ describes the mean gradient of heavy element mass fraction, i.e. the value of $\d Z/\d r$ averaged over several convective/diffusive cells, and $\delta_Z$ is the length over which compositional jump occurs. To constrain the value of $\delta_Z$, two limiting arguments can be used. On one side, equating the diffusive timescales for heat and solute leads to \balign{\frac{\delta_Z}{\deltat}\approx\sqrt{\frac{D}{\kapt}}. \label{dtdz_diff_limit}} On the other hand, considering that the compositional jump must cover the whole interface to ensure its mechanical stability, one would expect that \balign{\delta_Z\approx\deltat. \label{dtdz_stab_limit}}

The present paper focuses on the impact of double-difffusive convection on the planet mechanical and thermal {\it structure}. The impact on the evolution will be addressed in a forthcoming paper (see \sect{sec:evolution}). Therefore, this uncertainty on the heavy element flux, discussed in the next section, does not have a significant impact on the results presented in the present study. Determining more precisely the solute transport properties in the regime of layered convection, however, will be of central importance to evolutionary calculations, and 3D hydrodynamical simulations in a realistic parameter range are strongly needed \citep{MGS11}.

The term $\nabla_{\mu,I}$, which is the relevant quantity when computing the equilibrium condition of the interface, 
is given by
\balign{\label{dmu_dz}
\nabla_{\mu,I}=-\frac{l+\delta_Z}{\delta_Z}\hp \pdc{\ln \mu}{\,Z}{P,T} \langle\nabla Z\rangle.
}
The precise value of $\pdc{\ln \mu}{Z}{P,T}$ depends on the precise chemical composition of the considered heavy element, but is typically around unity; this is the value we will use in numerical estimates below. We will also use $D\approx 10^{-7}-10^{-8}$\,m$^2$.s$^{-1}$ (i.e. $\tau\approx10^{-2}-10^{-1}$), appropriate for giant planet interior conditions \citealt{SS77a}).


\section{Layer size: analytical point of view}\label{sec:mlt_limit}

\subsection{Existence of an equilibrium height}\label{sec:merging_equilibrium}

The determination of the size of the layers, when layered convection is occurring, is a complex task. The problem is rendered even more difficult by the fact that the small layers that initially form tend to merge into larger layers. The question is then to know \textit{if} and \textit{when} layer merging ends.

In the fingering case, i.e. when the solute is the destabilizing actor, \Citet{Rad05} derived a criterion for the linear stability of thermo-compositional staircases against merging. In this picture, an initially inhomogeneous medium starts forming layers if its density ratio ($R_\rho$) is smaller than the density ratio, $R_\mathrm{min}$, for which the heat to solute buoyancy flux ratio ($\equiv\gamma$) stops decreasing when $R_\rho$ increases. Then, layers are unstable and merge as long as the density ratio within the interface between them ($R_I$) is smaller than $R_\mathrm{min}$. Then, as $R_I$ increases with the layer height (under some conditions), it eventually reaches $R_\mathrm{min}$ and the merging process stops. At this stage, an \textit{equilibrium height} is reached by the convective layers and the staircase is fully equilibrated.

The linear stability analysis of \citet{Rad05} can be redone in the diffusive case, by simply accounting for the fact that the signs of the various density and temperature differences and of the fluxes must be changed. Then, after some lengthy but straightforward algebra, one can show that layers are unstable and merge when
\balign{\label{merge_crit} \pdc{\,\gamma^{-1}}{\,\Rzm}{\Rzm=\RIm}<0 \,\, \Leftrightarrow \,\,\RIm<\Rmin,}
which defines $\Rmin$. 
As for the fingering case, this linear stability analysis shows that, \textit{if layers form, their merging will stop when their height reaches a finite equilibrium value.}

\subsection{Global constraints}

Although precisely estimating this equilibrium value is difficult, some strong limits on the layer size can be derived theoretically.
 Let us consider a stack of layers extending over a zone of size $L$, and define a global gradient $\langle \nabla Z\rangle\approx -\Delta Z /L$, where $\Delta Z$ is the difference between the mass ratios of heavy element at the bottom and at the top of the semi-convective zone.

On one hand, the mean molecular weight gradient in \textit{all} the interfaces, $\nabla_{\mu,I}$,  must be high enough to satisfy the stability criterion discussed in sections \ref{sec:adiab} and \ref{sec:interfaces}. This implies
\balign{\label{stab}
\frac{\amu}{\at}\nabla_{\mu,I}> \Rmin \times(\deld-\delad).
}
 Substituting $\nabla_{\mu,I}$ by its expression in \eq{dmu_dz} and $\delta_Z$ by using \eqs{dtdz_diff_limit}{del_over_l}, the criterion (\ref{stab}) for a planetary scale ($L\approx\Rp$) semi-convective zone reads
\balign{
 \alpha^{1/(1+a)}> \alpha_\mathrm{min}^{1/(1+a)}\equiv \frac{\at}{\amu}\sqrt{\frac{D}{\kapt}}\frac{\Rmin (\deld-\delad)^{1-\frac{1}{4(1+a)}}}{( \Phi_0/\Cl)^{\frac{1}{4(1+a)}}\Delta Z\pdc{\ln \mu}{Z}{P,T}},
}
where $\alpha_\mathrm{min}$ denotes a lower limit for the layer size.
Under the present conditions in the interiors of our gas giants (see end of \sect{sec:conv_layers} and \tab{tab:constraints}), this yields $\alpha_\mathrm{\min}\approx 2 \times 10^{-9}\times (\Rmin)^{1+a}$, with the less restrictive constraint being obtained for $\Rmin=1$, as summarized in \tab{tab:constraints}. However, as the existence of layers allows $\Rmin$ to be as high as $(1+Pr)/(\tau+Pr)\approx 10^1-10^2$, these constraints could be severely tightened, as showed in \tab{tab:constraints}. 
This simply confirms that, \textit{in order for layered convection to be stable, convective cells must remain larger than the diffusive interfaces, and the medium is always in the convective regime, $\alpha\gg\alphac$.}

\begin{table}[htb]
\begin{center}
\caption{Numerical constraints on the layer height in the two limiting cases of convection ($a=0.2$ and 0.5; $\Cl=1$) for the following conditions (representative of Jupiter interior): $\Phi_0= 3\times 10^{33}$, $\surabd=\,10$, $\kapt=5\times 10^{-5}$\,m$^2$.s$^{-1}$, $D=5\times 10^{-7}$\,m$^2$.s$^{-1}$.}
\label{tab:constraints}
\small
\begin{tabular}{l l l r@{$\times$}l r@{$\times$}l } \hline\hline 
	& constraint &  &\multicolumn{2}{c}{ $a\,=\,0.2$} &\multicolumn{2}{c}{
$a\,=\,0.5$}  \\ \hline
$\alpha_\mathrm{min}$  & stability & $\Rmin=1$&2.4&$10^{-9}$  &  2.4&$10^{-9}$  \\
  &  & $\Rmin=\frac{1+Pr}{\tau+Pr}$ &6.0&$10^{-7}$  &  2.4&$10^{-6}$  \\
\hline
$\alpha_\mathrm{max}$ & homogeneization  & $Nu_\mu=1$& 1.0&$10^{-4}$  &  1.4&$10^{-3}$  \\
 &   & $Nu_\mu=3$ &2.5&$10^{-5}$  &  2.6&$10^{-4}$  \\
 \hline
 $\alpha_\mathrm{min}$ & observational & & \multicolumn{2}{c}{Jupiter} & \multicolumn{2}{c}{$3\times 10^{-5}$} \\
 & constraints$^\star$ & &\multicolumn{2}{c}{Saturn} & \multicolumn{2}{c}{$4\times 10^{-6}$}\\
\hline\hline
\end{tabular}
\end{center}
$^\star$ See \sect{sec:resnum}
\normalsize
\end{table}

On the other hand, the solute gradient within the planet will be homogenized within a typical timescale 
\balign{t_Z\approx\frac{\rho\,\Delta Z \Rp}{|F_Z|}.\label{tz}} 
Using Eq.\,(\ref{flux_Z}), (\ref{dmu_dz_def1}) and (\ref{dtdz_diff_limit}), and taking $\langle \nabla Z\rangle\approx \Delta Z/\Rp,$ \eq{tz}
  becomes
\balign{
t_Z \approx \frac{\Rp^2 }{D}\frac{\delta_Z}{l} \approx \frac{\Rp^2 }{\sqrt{\kapt\,D}}\,\left(\frac{\Phi}{\Cl}\right)^{-\frac{1}{4(1+a)}}\propto \alpha^{-1/(1+a)}. 
}
Therefore, in order to avoid a complete homogenization of giant planet interiors in less than 5\,Gyr, their present age, $\alpha$ must be smaller than $\alpha_\mathrm{max}\approx 1\times10^{-4}$ for $a=0.2$ and $1.4\times10^{-3}$ for $a=0.5$. Note however that, as mentioned in \citet{RGT11}, turbulent transport due to the double-diffusive instability can yield compositional Nusselt numbers, $Nu_\mu$\footnote{Analogously to the thermal Nusselt number, the flux of heavy elements, for a given $Nu_\mu$,  is given by $F_Z\equiv - \,Nu_\mu \times \rho\, D \,\nabla Z$.}, around 2-4, yielding even stronger constraints, as summarized in \tab{tab:constraints}. These small values of $\alpha$ justify a posteriori the
approximation of continuous thermal and heavy element profiles when considering the planet's entire internal structure. Note, however, that layered inhomogeneities could be dynamically regenerated over time. In that case, layered convection will be an ongoing process in the planet's interior.

Considering the possibility that the ratio of the compositional to the thermal size of the interface does \textit{not} scale as the square root of the ratio of the diffusivities, and therefore using \eq{dtdz_stab_limit} instead of \eq{dtdz_diff_limit}, yields an increase of both $\alpha_\mathrm{min}$ and $\alpha_\mathrm{max}$ by a factor $\sqrt{\kapt/D}^{1+a}\approx10$.

\textit{This analysis shows that, for the age of the Solar System and for the conditions prevailing in gas giant interiors, there exists a range of layer sizes for which ongoing layered convection is a viable mechanism.} According to our estimate, this range is relatively large and corresponds to :
\balign{
10^{-9}-10^{-6}\lesssim\alpha\lesssim10^{-4}-10^{-2}. 
}
The uncertainty on the lower and upper bounds are respectively due to our poor knowledge of the behavior of $\Rmin$, and of the solute flux at low Prandlt number. Even given these uncertainties, the fact that convection is always in a regime of relatively high Rayleigh number appears to be robust prediction.
To make an attempt to overcome these limitations, we examine in \sect{sec:resnum}  how observational data can narrow this possible domain of $\alpha$, by further constraining the degree of super adiabaticity, and thus the size of the layers, in our Solar System gas giants.

\subsection{An alternative scenario}

We stress that the aforementioned constraints apply only to \textit{layered} convection, and does not preclude the possibility that, under some conditions, double diffusive convection may manifest itself under the form of \textit{homogeneous} double diffusive convection and act like a \textit{turbulent diffusion}  \citep{RGT11}. However, for this to happen, the criterion (\ref{stab}) must be verified in the \textit{inefficient convection} regime, $\delta_T\gg l$ (see \app{sec:mlt_app}). The following criterion must then hold
\balign{
-\hp\langle\nabla Z\rangle \approx \Delta Z\,\frac{\Rp}{L}\, \gtrsim\,\Rmin \times(\deld-\delad).
}
As $\Delta Z\leqslant1$ and $\Rmin\geqslant1$, by definition, a zone of turbulent diffusion cannot extend over the {\it entire planet's scale} unless $\deld \lesssim~ \delad$, i.e. if the whole object is nearly diffusive in the first place. 

In an object with a heat flux high enough to be convectively unstable, the semi-convective zone must then be confined to a {\it fraction of the planet}, in particular, but not necessarily, near an immiscibility region or a phase transition for instance. In this case, the total size of the zone must verify $L/\Rp\leqslant \surabd^{-1}$ ($\sim 1/10$ in Jupiter), condition for which a large enough jump in the heavy element mass fraction can be sufficient to stabilize the whole zone against convection and open a \textit{diffusive buffer} in the interior (where $\delt\approx\deld$). From the global point of view of the planet, this would act as a composition, temperature and entropy nearly discontinuity.
This possibility for the existence of such a diffusive-like buffer in the interior of our gas giants, as a consequence of double-diffusive instability, must be kept in mind.

In the following, however, we will not consider this scenario any further, and we will only consider the effect of a planetary scale layered-convection zone.


\section{Numerical results for Solar System giant planets}\label{sec:resnum}

In this section, we examine whether the presence of semi-convection in Jupiter and Saturn interiors can be consistent with the various available observational constraints.
We first derive homogeneous reference interior models in \sect{sec:refcase}. Then, in \sect{sec:numres}, 
 we incorporate our model for layered convection into the standard method used to compute interior structure models of rotating gaseous planets (presented in \sect{sec:method}) and determine the area of the composition/layer size space parameter which is consistent with observed gravitational moments and surface abundances.

\subsection{Hydrostatic equilibrium and figures of the planet}
\label{sec:method}

Solar System giant planets are rapidly rotating bodies (the period of rotation is about 10 hours), with the centrifugal potential representing about 10\% of the gravitational potential. This modifies the hydrostatic equilibrium condition between the pressure gradient and the gravitational force in the interior, which now writes
\balign{ \label{HSE}
\nabla P=-\,\rho \,\nabla(\Vg+\Vrot),
}
where 
\balign{
\Vg(\mathbf{r})=-\,G\int \frac{\rho(\mathbf{r}^\prime) }{ |\mathbf{r}-\mathbf{r}^\prime |} \, \d^3\mathbf{r}^\prime
}
and
\balign{
\Vrot(r,\theta)=-\int_0^\xi \ind{\omega}{p}^2(\xi^\prime)\,\xi^\prime\d\,\xi^\prime
}
denote respectively the gravitational and centrifugal potentials, 
with differential rotation $\ind{\omega}{p}(\xi)$, where $\xi$ is the distance from the position $\mathbf{r}$ to the rotation axis, and $G$ the gravitational constant. In the present study, $\ind{\omega}{p}$ is assumed to be constant and given by the magnetospheric rotation rate.
Because of the symmetry of the centrifugal potential with respect to both the rotation axis and the equatorial plane, surfaces of equal densities for these objects are supposed to be generalized ellipsoids of revolution whose exact shape is given by
\begin{align}\label{def_rb}
r(\bar{r},\theta)=\bar{r}\,\left[1+\sum_{\n} s_{2\n}(\bar{r})\,P_{2\n}(\cos \theta)\right],
\end{align}
where $\bar{r}$ is the mean radius of the equipotential, $P_{2\n}$ are the usual Legendre polynomials, $\theta$ is the colatitude and the $s_{2\n}$ are a set of figure functions. These latter can be derived using the theory of figures for rotating bodies \citep{ZT78}, and must be solved iteratively with the set of \textit{perturbed} 1D hydrostatic equilibrium equations
\balign{
\pd{\,P}{\,m}&=-\frac{1}{4\pi}\frac{ G m}{ \bar{r}^4}+\frac{ \op^2}{6\pi \bar{r}}+\varphi_\omega(\bar{r}), \\
\pd{\,\bar{r}}{\,m}&=\frac{1}{4\pi \bar{r}^2\rho}, \\
\pd{\,T}{\,m}&= \frac{T}{P}\pd{\,P}{\,m}\delt, \label{chess:dtdm} 
}
where $m$ is the mass enclosed in the equipotential of mean radius $\bar{r}$, $\op$ is the rotation rate of the planet, $\varphi_\omega(\bar{r})$ is a second order correction due to the centrifugal potential, which depends on the figure functions. As discussed in \sect{sec:mlt}, the prescription to be used for $\delt$ is determined by the energy transport processes. 

 The departure from sphericity of the iso-density surfaces results in a perturbation of the external gravity field $\Vg(r,\theta)$ that writes
\begin{align}
\Vg(r,\theta) &= -\frac{GM}{r} \Bigl\{ 1-\sum_{\n=1}^\infty \left(\frac{\Req}{r}\right)^{2\n} J_{2\n}P_{2\n}(\cos \theta) \Bigr\},\\
J_{2\n} &= -\frac{1}{M\Req^{2\n}} \int_V \rho(r,\theta)\,r^{2\n}P_{2\n}(\cos \theta)\,\d^3\mathbf{r},
\end{align}
\noindent where
$r$ is the radial distance from the center of the planet, $M$ the mass of the planet, $\Req$ the equatorial radius, $\theta$ the colatitude, $P_{2\n}$ are Legendre polynomials of order $2\n$ and
$J_{2\n}$ denote the gravitational moments, that can be computed once the figure equations have been solved.
 The measured gravity moments provide stringent constraints on the
density profile and the possible layering within these planets.

As, in practice, Legendre polynomial expansions are truncated at a given order $\n$, a closure equation is provided by the equation of state (EOS) of the mixture along the planet's interior profile. Such an EOS is generally given by the so-called ideal volume law for the mixture:
\balign{
\frac{1}{\rho}=\frac{X}{\rho_X}+\frac{Y}{\rho_Y}+\frac{Z}{\rho_Z},
}
where $X$, $Y$ and $Z$ denote the mass fractions of H, He and heavy elements, respectively.
For the H/He fluid, the most widely used EOS
is the Saumon-Chabrier-vanHorn EOS (\citealt{SCV95}; SCvH). For the heavy material, we have used the "Rock" EOS of \citet{HM89} for silicates and the "Ice" ANEOS equation of state \citep{TL72} for volatiles (CH$_4$, NH$_3$, H$_2$O). The impact of the differences between various EOS's on exoplanet structure and evolution has been explored in \citet{BCB08}.

Once such EOS's, $P[\rho(X_i)]$, are specified,
structure models with various compositions are calculated by solving iteratively the aforementioned hydrostatic equilibrium condition for a rotating body and the third-order level-surface theory \citep{ZT78} to obtain a model which reproduces the observed values of the radius, $\Req$, and gravitational moments $J_2$ and $J_4$ measured by the Pioneer and Voyager missions (see Table \ref{tab:data}).

\begin{table}[htb]
\begin{center}
\caption{Observed characteristics of Solar System gaseous giants (\citealt{Gui05} and references therein; the numbers in parentheses are the uncertainty in the last digits 
of the value).}
\label{tab:data}
\small
\begin{tabular}{l r@{.}l r@{.}l } \hline\hline 
	&\multicolumn{2}{c}{\bf Jupiter} &\multicolumn{2}{c}{\bf
Saturn}  \\ \hline
$\Mp$ [$10^{26}$kg] & 18&986112(15)  & 5&684640(30)  \\
$R_{\rm eq}$ [$10^7$m] &  7&1492(4)  & 6&0268(4)  \\
$R_{\rm pol}$[$10^7$m] &  6&6854(10)  & 5&4364(10)    \\
$P_{\mathrm{rot}}\,$[$10^4$s] & 3&57297(41)  & 3&83577(47)  
\vspace{3pt}\\
$T_{\mathrm{1bar}}\,$[K] & 165&(5)  & 135&(5) \\
$\ftot\,$[W.m$^{-2}$] & 5&44(43)  & 2&01(14) 
\vspace{3pt}\\
$J_2\times10^2$ & 1&4697(1)  & 1&6332(10)    \\
$J_4\times10^4$ & -5&84(5)  & -9&19(40)    
\vspace{3pt}\\
$Z_\mathrm{atm}/Z_\odot$ & \multicolumn{2}{c}{2-4} & \multicolumn{2}{c}{2-8} \\
$(Y/(X+Y))_\mathrm{atm}$ & 0&238(50) & 0&215(35) \\
\hline\hline
\end{tabular}
\normalsize
\end{center}
\end{table}

\subsection {Reference homogeneous models}
\label{sec:refcase}

In conventional giant planet models, the abundances of heavy elements are chosen to be constant in the gaseous H/He envelope, with a possible discontinuity at the transition between the molecular and metallic regions \citep{CSH92}. Under the actual conditions found in Jupiter and Saturn, the thermal gradient that would be needed to transport the whole flux by diffusive processes, $\deld$, is always larger than $\delad$ and the whole interior is convective in the homogeneous case according to the Schwarzschild criterion. As convection is very efficient (see \sect{sec:mlt}), the super adiabaticity needed to transport the outgoing energy is on the order of $10^{-8}-10^{-9}$, so that the structure can be solved by setting $\delt=\delad$ in \eq{chess:dtdm}.

\begin{figure}[tbp] 
 \centering
 \resizebox{.9\hsize}{!}{\includegraphics{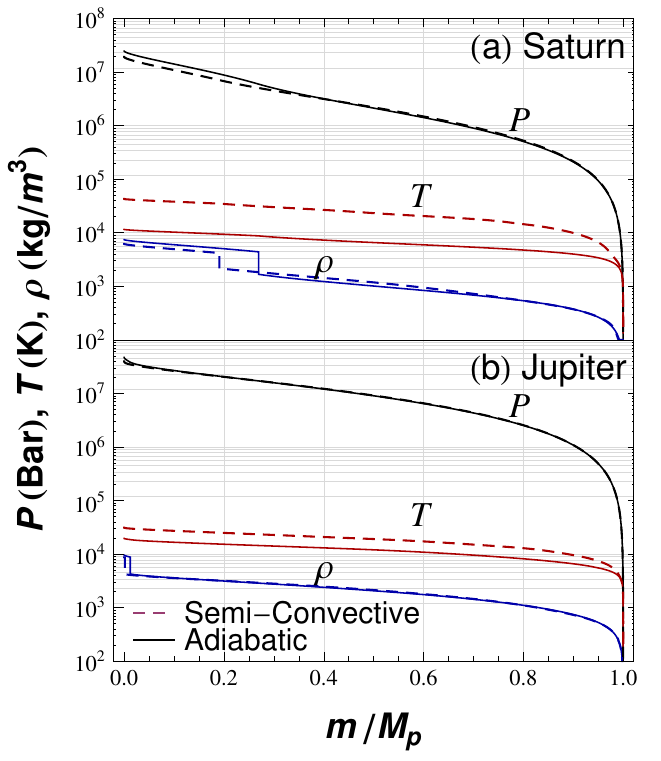}}
 \caption{
Pressure (black), temperature (red) and density (blue) profiles as a function of depth (expressed by the Lagrangian coordinate, i.e. the mass $m$), for the reference adiabatic (solid curves) and semi-convective (dashed curves) cases for  Saturn (a) and Jupiter (b). The increased thermal gradient due to the inefficient heat transport in the semi-convective case (with $\NL=10^4$ for Jupiter and $10^{4.5}$ for Saturn) strongly increases the internal temperature. This causes a partial redistribution of the core material within the gaseous envelope.}
 \label{fig:profiles}
\end{figure}

In order to have a reference case, we use the formalism described in \sect{sec:method} to derive \textit{homogeneous}, adiabatic interior models representative of the usual 2-layer composition. As we use the interpolated SCvH EOS, we do not consider the effect of a Plasma Phase Transition, and we are thus left with only two free parameters, namely the core mass ($\mcore$) and the metal mass fraction in the gaseous envelope ($\Zenv$)\footnote{As already found by \citet{CSH92} and \citet{SG04}, with the SCvH EOS, simple homogeneous models such as our reference case cannot reproduce $J_4$ to better than a few percents error.}. The temperature, density and pressure profiles of our best representative homogeneous models of Jupiter and Saturn are shown in \fig{fig:profiles} (solid curves). These are composed of a solid core of mass $\mcore=3.9$ and $25.6\,\mearth$ surrounded by a H/He gaseous envelope with a {\it constant}
metal fraction $\Zenv=0.11$ and $\Zenv=0.05$ for Jupiter and Saturn, respectively (these results are summarized in Table \ref{tab:res}). 
These reference models yield interior enrichment that are consistent with previous determinations \citep{CSH92,SG04,Gui05}.

%



\subsection{Inhomogeneous models}\label{sec:numres}

We now derive semi-convective, inhomogeneous interior models for Jupiter and Saturn. We stress that
\textit{all these models are consistent, within the observational uncertainties,  with the measured gravitational moments of Jupiter and Saturn 
}  (see Table~\ref{tab:data}; \citealt{CS85}; \citealt{CA89}).

An additional constraint on the outermost value of the compositional gradient is provided by the surface abundance of heavy elements in the planets measured by the 1995 Galileo Entry Probe mission. Indeed, elemental abundances of the atmospheres of solar giant planets are observed to differ significantly from each other and from the solar composition, being enriched by a factor $\sim 2-4$ and $\sim2-8$ with respect to the Sun's atmosphere for Jupiter and Saturn,  respectively, as shown in Table \ref{tab:data} \citep{Gui05}. 
Moreover, the planet's total mean abundances of H and He ($\bar{X}$ and $\bar{Y}$) must recover the values of the protosolar nebula, i.e. $\bar{Y}/(\bar{X}+\bar{Y})\approx 0.275$. 

In our calculations, the adjustable parameters to fulfill all these constraints are chosen to be the mass of the core ($\mcore$), the mean heavy element mass fraction in the gaseous envelope (${\bar Z_\mathrm{env}}$), and the global compositional variation in the envelope ($\Delta Z_\mathrm{env}$, the difference between the metal mass fraction just above the central core and the one in the atmosphere).
To assess the robustness of our results with respect to the equation of state chosen to describe the thermodynamics of the heavy material, we derived several sets of models for which the composition of the core varies from pure ice to pure rock.

\subsubsection{Number of layers}

From a macroscopic point of view, an important quantity describing layered convection is the number of convective-diffusive layers, $\NL$. This number is roughly equal to the ratio of the size of the semi-convective zone, comparable to the
planet's radius, $\Rp$, if this zone extends over the whole planet, to the height of a
typical convective/diffusive cell, $l+\delta_T$.
As shown in \sect{sec:mlt_limit}, 
in the regime of interest, $l+\delta_T\approx l$, and $\NL\approx{\Rp/l}$.
Because $\hp\approx \Rp$ in the deep interior, the number of layers in the planet is thus approximately equal to $\NL\sim \alpha^{-1}$, and in the following we will always refer indifferently to either $\alpha$ or \balign{\NL\equiv\alpha^{-1}\equiv\hp/l.}

\begin{figure}[tbp] 
 \centering
 \resizebox{.9\hsize}{!}{\includegraphics{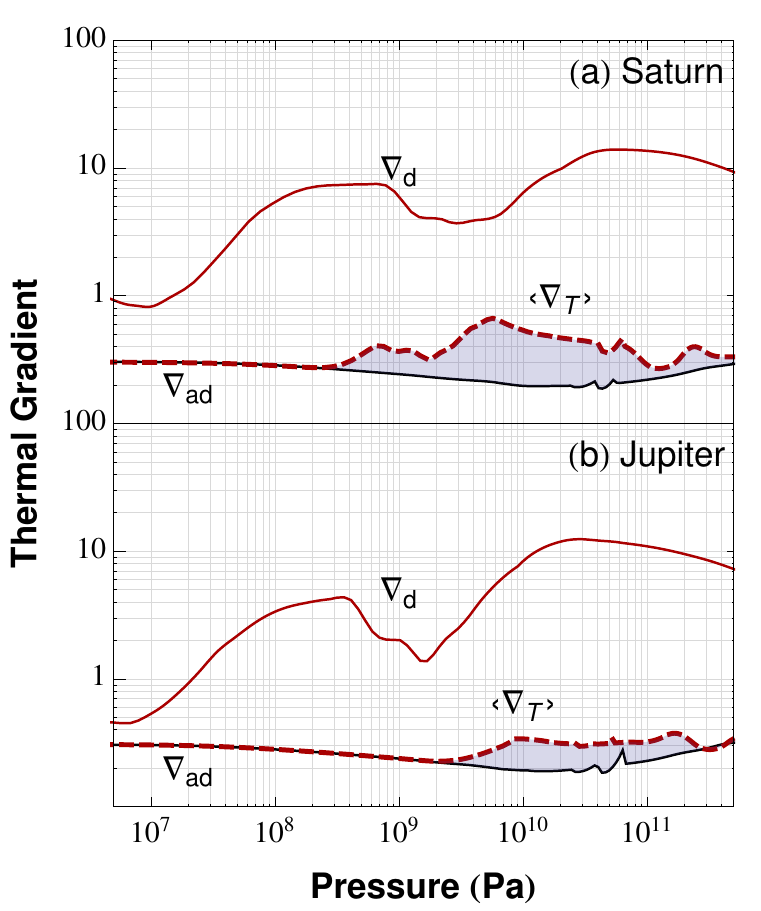}}
 \caption{
Conductive (top red curves) and adiabatic (bottom black curves) thermal gradient profiles for Jupiter and Saturn. The dashed curves correspond to the mean thermal gradient for models with $10^{4.5}$ and $10^{5.4}$ layers for Jupiter and Saturn, respectively. The shaded area represents the allowed range of super adiabaticity in presence of semi-convection,
consistent with the observational constraints. }
 \label{fig:thermalgradient}
\end{figure}
%

%
%

As shown in \sect{sec:mlt_limit}, $\NL$ is constrained to lie
within the range $10^{2-4}\le\NL\le 10^{6-9}$. Note that, given the small size of the diffusive-convective layers compared with the size of the planet, the discontinuous (staircase-like)
temperature and composition profiles can be well approximated by continuous \textit{mean} thermal and compositional gradients ($\delmean$ and $\langle\delmu\rangle$, respectively) to determine the planet's global structure, as illustrated in \fig{fig:cartoon}.

This possible range of numbers of layers is further constrained by our numerical calculations,
which show that, in order to reproduce our giant planet observational constraints, no more than $\sim 2.5\times 10^5$ layers ($\alpha_\mathrm{min}\approx4\times10^{-6}$) can in reality be present in Saturn and $\sim 3\times10^{4}$ ($\alpha_\mathrm{min}3\times10^{-5}$) in Jupiter (see \tab{tab:constraints}). Indeed, a larger number of layers leads to so high temperatures in the interior
that the induced mean density decrease can not be counterbalanced by an increase of the heavy element mass fraction compatible with the observed
surface abundances. 
This is due to the fact that, the larger the number of layers, the smaller the size of each convective cell, reducing the maximum height a convective eddy can travel to transport heat before being stopped by the negative buoyancy present in the diffusive interface. 

A large number of layers thus decreases convective heat (and composition) transport efficiency. This leads to an increase of the mean super adiabaticity, as portrayed on \fig{fig:thermalgradient}, which in turn immediately implies a rise of the internal temperature, as illustrated on \fig{fig:profiles}. It is important to stress that super adiabaticity is the physical quantity most directly constrained by the data. Thus, \textit{whereas the allowed range of number of layers (or equivalently of layer sizes) may \textit{slightly} depend on the model used to parametrize semi-convection (see \sect{sec:mlt}), the allowed range of super adiabaticity displayed in \fig{fig:thermalgradient} should remain weakly affected}.

\subsubsection{Enhanced heavy material enrichment}

The pressure, density and thermal profiles obtained in the most extreme \textit{semi-convective} case compatible with the observational constraints discussed above are shown in \fig{fig:profiles} (dashed curves). As seen on the figure, and as expected from the above discussion, the non-adiabatic envelope profile obtained in the semi-convective case yields substantially higher internal temperatures than the usual adiabatic assumption, as heat and material redistributions are partly inhibited by diffusive processes. The pressure and density profiles, on the other hand,
remain barely affected, being strongly constrained by the gravitational moments.
\begin{figure}[tbp] 
 \centering
 \resizebox{1.\hsize}{!}{\includegraphics{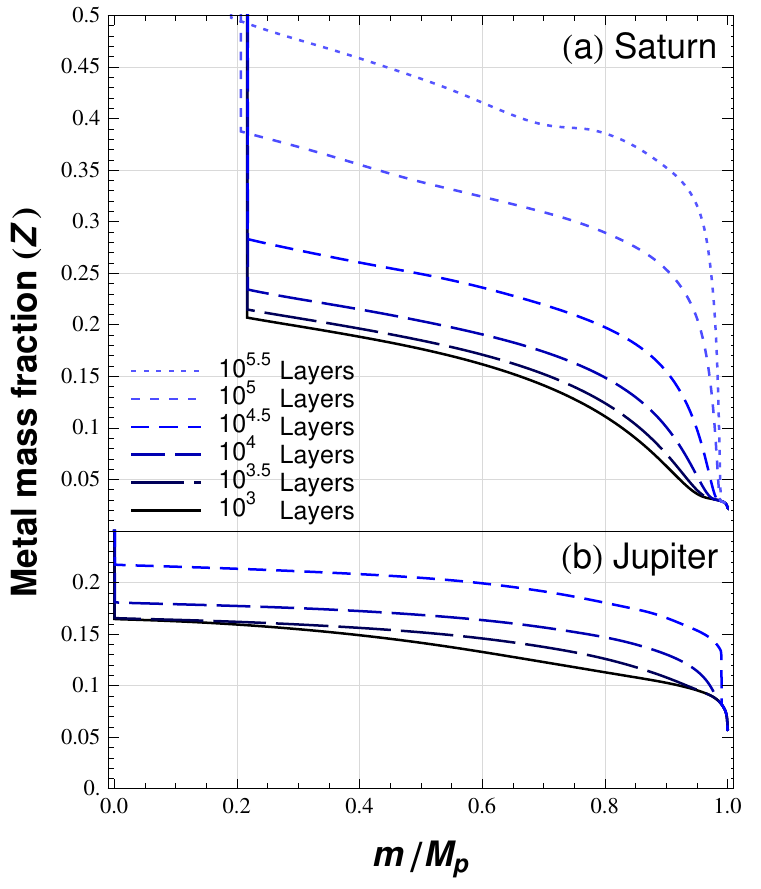}}
 \caption{
Metal abundance profiles as a function of depth (expressed by the Lagrangian mass $m$) for  Saturn (a) and Jupiter (b) for different numbers of layers. The abundance increases with the number of layers, to keep the density profile unchanged when convection becomes inefficient. The extreme cases ($10^{5.4}$ layers for Saturn and $10^{4.5}$ layers for Jupiter) correspond to the semi-convective profiles portrayed in \fig{fig:profiles}.}
 \label{fig:ZYprofiles}
\end{figure}
Hence, at basically fixed density profile, a higher temperature profile must be compensated by a larger amount of heavy material within the envelope.
This is illustrated in \fig{fig:ZYprofiles}, where we show the abundance profiles, as calculated in \app{sec:HEcontent_calc}, corresponding to semi-convective models with different numbers
of layers. The bottom curve (solid) in each panel corresponds to models with 1000 layers while the other curves correspond to a gradually increasing
number of layers.

\begin{figure}[htbp] 
 \centering
 \resizebox{1.\hsize}{!}{\includegraphics{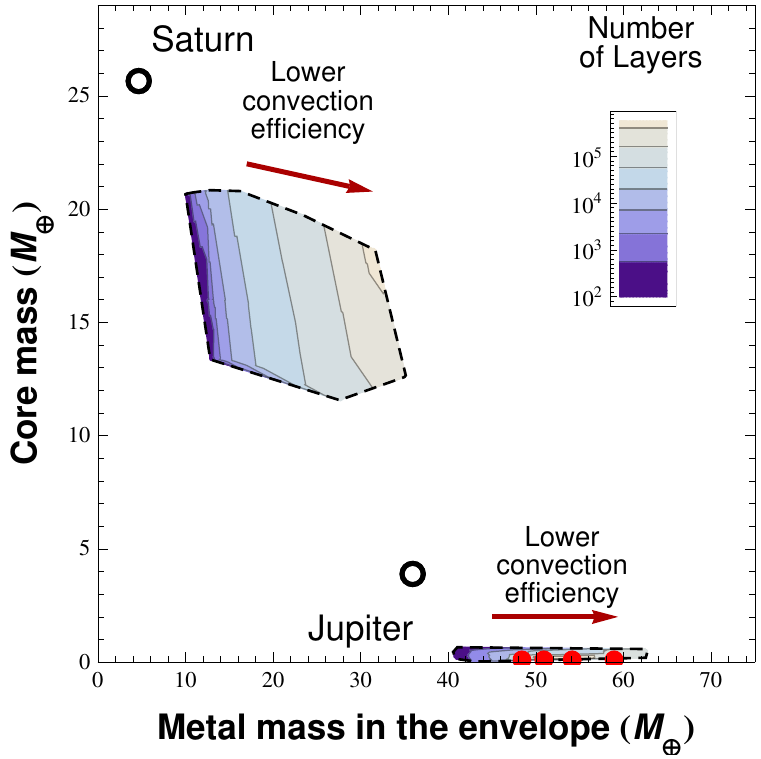}}
 \caption{Mass range of heavy elements in the core ($\mcore$) and in the envelope ($M_{Z,\mathrm{env}}$) consistent with all observational constraints, for different numbers
 of layers, for Jupiter (bottom right) and Saturn (upper left). The open dots at the upper left of each region correspond to the homogeneous interior models. As the number of semi-convective layers increases, the efficiency of convection decreases, and the heavy element mass fraction increases to counteract 
 the radius increase induced by the planet's higher internal temperature. The metals initially present in the core are then redistributed within the envelope. For Jupiter, solutions with no core at all ($\mcore=0$) can be found for the non adiabatic models (red dots).
}
 \label{fig:composition}
\end{figure}

Therefore, in order to compensate the radius increase (density decrease) due to the hotter interior, semi-convection yields a significantly larger \textit{total metal content} compared with conventional homogeneous models. This can be seen in \fig{fig:composition}, which shows the amount of heavy elements in the core and envelope for the various cases discussed here, as summarized in Table \ref{tab:res}. For Saturn, up to $50\,\mearth$ of heavy elements could be present in the planet while for Jupiter the heavy material content could reach 63 $\mearth$. This corresponds to about 25 and 10 times the solar abundances, respectively\footnote{ Note that the abundances of heavy elements brought to Jupiter and Saturn, in particular water, could already be enriched compared with the solar value \citep{GHM01}}. Since these values only depend on the allowed amount of super adiabaticity, they should not strongly depend on our modeling of diffusive/convective transport, as mentioned above.
In contrast, the maximum amount of heavy elements compatible with the observational constraints for the homogeneous, adiabatic models, is about $30\,\mearth$ for Saturn and 40 $\mearth$ for Jupiter, in agreement with previous studies \citep{SG04}.

But semi-convection does not only increase the global metal content, it also yields a {\it completely different distribution of heavy elements}.
While the global enrichment of the planet is increased
in the inhomogeneous models, the mass of the central core is decreased, as heavy elements are preferentially redistributed in the gaseous envelope.
In the case of Saturn, the vertical spread in core mass at fixed number of layers illustrated in \fig{fig:composition} is obtained when varying the core composition from pure ice (top) to pure rock (bottom). In Jupiter the inferred core mass is too small for the equation of state to make a significant difference. One could wonder why the homogeneous case is not continuously recovered when $\alpha$ tends toward 1. This slightly counter intuitive effect is due to the fact that, at least when using the SCvH EOS, completely homogeneous models (central core plus a fully homogeneous envelope) cannot in general reproduce both the observed $J_2$ and $J_4$ \citep{CSH92,SG04}. Thus, if we relax the constant $Z$ condition in the envelope, the presence of a compositional gradient and of a smaller core appears to be the best solution to reproduce observational data, even in the absence of any additional super-adiabaticity. 

 \textit{For Jupiter, models can be found that match the gravitational moments without the presence of a central, completely differentiated core (red dots on the bottom right of \fig{fig:composition})}. Such cases yield an atmospheric metallicity $\ind{Z}{atm}\sim4-5\,Z_\odot$. The fact that the possible erosion of the core mass would have been more efficient in Jupiter than in Saturn might stem from
the larger energy flux available in Jupiter \citep{GSH04}.

\begin{table}[htb]
\begin{center}
\caption{Heavy element content for Jupiter and Saturn inferred from the various models consistent with these constraints within the quoted observational uncertainties.}
\label{tab:res}
\small
\begin{tabular}{l r@{.}l r@{.}l } \hline\hline 
	&\multicolumn{2}{c}{\bf Jupiter} &\multicolumn{2}{c}{\bf
Saturn}  \\ \hline
Region&\multicolumn{4}{c}{Amount of heavy elements ($\mearth$)}
\vspace{3pt}\\
&\multicolumn{4}{c}{Homogeneous model}\\
Envelope & \multicolumn{2}{c}{36} & \multicolumn{2}{c}{4.7} \\
Core & \multicolumn{2}{c}{3.9} & \multicolumn{2}{c}{25.6} \\
Total & \multicolumn{2}{c}{40} & \multicolumn{2}{c}{30.3} \\
&\multicolumn{4}{c}{Semi-convective models}\\
Envelope & \multicolumn{2}{c}{41-63.5} & \multicolumn{2}{c}{10-36} \\
Core& \multicolumn{2}{c}{0-0.5} & \multicolumn{2}{c}{10-21}\\
Total & \multicolumn{2}{c}{41-63} & \multicolumn{2}{c}{26-50} \\
\hline\hline
\end{tabular}
\normalsize
\end{center}
\end{table}

\section{Prospect for giant planet evolution}\label{sec:evolution}

While the aim of present study is to examine and to constrain the properties of transport by semi-convection inside Jupiter and Saturn \textit{at the present time}, the impact of non-adiabatic interiors on the cooling of these planets, and of giant planets in general, remains to be explored. This requires more cumbersome evolutionary calculations, for which initial conditions will be of crucial importance, as will be explored in a forthcoming study. Note that evolution will add an additional constraint, namely that the planet cooling rates yield the correct properties at the age of the solar system, which will put more stringent constaints on the range of possible layer sizes/numbers. Without going into such detailed calculations, however, the following points can be mentioned.

\subsection{Merging of the layers}

As mentioned in \sect{sec:mlt_limit}, soon after they form, layers are expected to merge, leading to larger layers, until the layer height reaches either the planet size, yielding a standard adiabatic interior, or an equilibrium value, which is the semi-convective case considered here. Numerical simulations by \citet{Rad05} and \citet{RGT11} tend to show that the equilibration timescale of the staircase is much shorter than the typical timescale for the evolution of the planet.

Thus, if an equilibrium height is reached, as discussed in \sect{sec:merging_equilibrium}, this rather quick equilibration timescale suggests that the layer size should remain roughly constant during the evolution, or change slowly with the mean properties of the medium (e.g. the mean molecular weight gradient). A precise prescription for the height of the layers being yet lacking and demanding a more precise knowledge of the heat transport properties of layered convection under astrophysical conditions, it seems reasonable, as a first guess, to use a constant size, whose value has been constrained in the afore sections, throughout the evolution.

\subsection{Initial heavy element distribution}

Conventional models based on fully adiabatic thermal profiles notably lead to cooling times about 15\% longer than the age of the Solar System for Jupiter \citep{FIN11}.
In principle, the hotter non-adiabatic internal structures suggested in the present paper will prolong the cooling and thus worsen the problem.

However, in the case of the erosion of an initially large core, part of the gravitational work will be spent eroding the core and mixing the material upward and will thus not contribute to the total luminosity, quickening the cooling.  
All these effects must be properly accounted for to infer the appropriate cooling timescale.

In addition, if Jupiter and Saturn initial cores were allowed to be relatively large ($\gtrsim 10\,\mearth$), the corresponding high surface density of solids in the protosolar nebula will quicken the formation timescale in the conventional core accretion scenario, helping solving the related
formation timescale problem \citep{PHB96}.
 Finally, since, in the present scenario,  some of the ablated material from the accreted planetesimals during the planet's early formation stages
remains distributed throughout the envelope, this will (i) reduce the heating due to gravitational energy release
produced by the infalling planetesimals on the planet embryo and (ii) increase the envelope mean molecular weight. Both effects will cause the protoplanet to contract more quickly, shortening again the planet's formation timescale in the conventional core accretion scenario \citep{PHB96}. A correct exploration of the impact of inhomogeneous interiors upon giant planet history thus necessitates to investigate the consequences not only
on the thermal evolution but also on the formation process.
 


\section {Conclusion and perspective}\label{sec:conc}

In this paper, we have first developed an analytical approach of layered convection, based on a standard parametrization of convection similar to the MLT formalism. This formalism allows a quantitative determination of the expected number of diffusive layers, or equivalently of the average characteristic mixing-length
parameter,  in a semi-convective planet interior characterized by a given total flux and a given thermal (and compositional) diffusivity. Furthermore, this formalism allows an exact determination of the characteristic thermal gradient in the presence of double-diffusive convection, and thus of the related amount of super-adiabaticity within the planet's interior.

Using this formalism, we have computed semi-convective interior models of Jupiter and Saturn.
We have shown that a stratified internal structure for Solar System gaseous giants, with a compositional gradient of heavy material extending over a
substantial fraction of the planet, is a viable hypothesis, as such models can fulfill all the observed gravitational and atmospheric constraints for these planets. 
This new possibility differs from the conventional description of giant planet interiors, assumed to be composed of 2 main superposed, well identified layers of
homogeneously distributed material, namely a solid core surrounded by a dominantly gaseous H/He envelope. The consequences of the present giant planet interior description
are multiple. Namely, 
\begin{itemize}
\item (i) our jovian planets might be significantly more enriched in heavy elements than previously thought, 
\item (ii) their interior temperature, thus heat content,
might be much larger than usually assumed, 
\item (iii) the inner temperature profile could significantly depart from the usually assumed adiabatic profile. 
\end{itemize}
We stress that these conclusions do not depend on the precise model used to describe double diffusive convection.
Besides directly affecting our conventional vision of giant planet mechanical, compositional and thermal structures, these results
have profound impacts on our understanding of planet formation and cooling properties. Indeed, the revised possible maximum amount of heavy material bears direct consequences on the determination of the efficiency of solid planetesimal accretion during planet formation in
the protoplanetary  nebula, suggesting an early and efficient capture of planetesimals for our, and probably extrasolar as well, giant planets. Moreover,
the larger heat content and the departure from adiabaticity, as well as the possibility of significant core erosion from an initially
large core, directly impact the planet cooling
histories. Departure from adiabaticity, in particular, implies less efficient heat transport, a direct
consequence of the inhibited convective motions due to a persistent compositional gradient, and thus a smaller heat flux output rate than assumed in the conventional approach.

These results open a new window, and raise new challenges, on our present understanding of planet structure, formation and evolution. Importantly, the viability of such stratified interior models for our Solar System gas giants
directly applies to the case of extrasolar planets, reinforcing the possibility that such a lower heat flux output could at least partly explain the anomalously large radius of several 
transiting "hot Jupiters" \citep{CB07}. Indeed, it seems that invoking an extra source of (tidal, kinetic or magnetic) energy dissipation
 in these object interiors can not completely solve this ``radius anomaly'' puzzle
 and that an alternative or complementary process is necessary \citep{LCA11}. Unconventional, inhomogeneous non-adiabatic planetary interiors, as suggested in the present study, might provide the missing piece of the puzzle.

\begin{acknowledgements}
We would like to thank our referee D. J. Stevenson for his sharp questions which considerably enlarged the scope of this study.
The research leading to these results has received funding from the European Research Council under the European Community's Seventh Framework Programme (FP7/2007-2013 Grant Agreement no. 247060)
\end{acknowledgements}

\bibliography{biblio} 
\bibliographystyle{aa}


\appendix

\section{The case of the mixing length theory}\label{sec:mlt_app}

Here, we briefly discuss the particular case of the Mixing Length Theory (MLT; \citealt{HK94}). This case can be recovered in the efficient convection regime by the more general model presented in \sect{sec:mlt}, by setting $\Cl=1$ and $a=1/2$, but we show below that the MLT formalism can be extended to the inefficient convection regime.

\citet{HK94} showed that in the $Pr\ll 1$ regime, recasting their Eq.\,(5.60) in the notation of \sect{sec:mlt}, the modified Rayleigh number verifies 
\balign{
 \label{mltequ}
\Phi\equiv \NuT\times \Ras=\Ras+\left(\Ras^{1/2}\cdot\sigma(\Ras)\right)^3,
}
with
\balign{\label{def_XiLambda}
\sigma(\Ras)&=\frac{1}{2\sqrt{\Ras}}\left(\sqrt{1+4\,\Ras}-1\right),
}
where $\sigma\,\Nt$ is the growth rate of a convective eddy and the inverse of the convective time. In general, these equations can be solved numerically. It can be easily verified that in the limit of efficient convection, i.e. $\Ras\gg1$, $\sigma\rightarrow 1$, meaning that radiative losses are negligible and that the convective time tends toward $\Nt$. In this limit, a simple expansion of \eq{mltequ} yields $\Phi\approx\Ras^{3/2}$ and $\NuT\approx\Ras^{1/2}$ as expected in the standard MLT formalism \citep{HK94}.

In the $1\gg\Ras\gg Pr$ regime, however, developing \eq{def_XiLambda} yields $\sigma\approx\Ras^{1/2}.$ Then, because the convective time is dramatically increased, \eq{del_over_l} rewrites in that case
\balign{\left(\fracl{\deltat}{l}\right) \approx\left(\fracl{\kapt}{l^2 \,\sigma \,\Nt}\right)^{-1/2}\approx \Ras^{-1/2}\gg 1.} The size of the diffusive interfaces thus grows until the convective layers eventually turn into a completely diffusive medium for which $\delmean=\deld$.



\section{Computation of the heavy element content}\label{sec:HEcontent_calc}

The mean molecular weight gradient needed to stabilize the fluid against large scale convection can be caused by an inhomogeneous distribution of both helium ($Y(m)$) and metals ($Z(m)$) in the hydrogen ($X(m)$)-rich medium. In practice, both gradients can be present at the same time and either compete or contribute constructively.

In our model, we consider an ideal mixture of heavy elements within a H/He envelope whose H/He mass ratio is kept constant and equal to its value in the protosolar nebula, (H/He)$_\mathrm{proto}$. This implies
\balign{
X+Y+Z=1,~\mathrm{and} ~ \frac{Y}{X+Y}=\left(\frac{\bar{Y}}{\bar{X}+\bar{Y}}\right)_\mathrm{proto}\approx 0.275,
}
everywhere in the planet's gaseous envelope. We are then left with only one degree of freedom. Following previous calculations \citep{CSH92}, for sake
of simplicity and in order to have a flexible determination of the metal enrichment and a thermodynamically consistent EOS in the gaseous phase, we approximate the metal mass
fraction by an \textit{effective} helium mass fraction ($Y'$) in the H/He EOS. For the core, the metal mass fraction
is correctly described by the appropriate water and silicate EOS mentioned in the text.

The various element mass fractions, then the corresponding metal enrichment, are thus inferred from the relation
\balign{
\frac{1}{\rho(P,T,Y')}=\frac{1-Z}{\rho_{\mathrm{(H/He)_{proto}}}(P,T)}+\frac{Z}{\rho_Z(P,T)},
}
which gives $Z$ at each depth along a given model $P $-$ T$ profile \citep{CSH92}. The hydrogen and helium mass fractions are then derived using
\balign{
Y=(1-Z)\left(\frac{\bar{Y}}{\bar{X}+\bar{Y}}\right)_\mathrm{proto}}
and
\balign{
X=1-Y-Z.
}
In this simple model, a $Z$ gradient thus necessarily yields a competing inhomogeneous helium distribution within the planet. Because the mean molecular weight of a H/He mixture at fixed temperature and pressure only depends on $Y/(X+Y)$, only the $Z$ variations need to be considered to compute $\delmu$ in our simplified model. 
The $Z(m)$ profile is then integrated to obtain the total amount of heavy elements mixed in the gaseous layers for each planetary model, as portrayed in \fig{fig:composition}. 
In the most general case, with an intrinsic inhomogeneity of the helium distribution, caused for instance by its immiscibility in metallic hydrogen, both the $Y$ and $Z$ gradients would have to be properly calculated.

\end{document}